# Applications of machine Learning to improve the efficiency and range of microbial biosynthesis: a review of state-of-art techniques

Akshay Bhalla[1], Suraj Rajendran[2]
[1]*Dhirubhai Ambani International School, Mumbai, Maharashtra, India*
[2]*Institute for Computational Biomedicine, Department of Physiology and Biophysics, Weill Cornell Medicine of Cornell University, New York, NY, USA*

**Abstract**
In the modern world, technology is at its peak. Different avenues in programming and technology have been explored for data analysis, automation, and robotics. Machine learning is key to optimize data analysis, make accurate predictions, and hasten/improve existing functions. Thus, presently, the field of machine learning in artificial intelligence is being developed and its uses in varying fields are being explored. One field in which its uses stand out is that of microbial biosynthesis. In this paper, a comprehensive overview of the differing machine learning programs used in biosynthesis is provided, alongside brief descriptions of the fields of machine learning and microbial biosynthesis separately. This information includes past trends, modern developments, future improvements, explanations of processes, and current problems they face. Thus, this paper's main contribution is to distill developments in, and provide a holistic explanation of, 2 key fields and their applicability to improve industry/research. It also highlights challenges and research directions, acting to instigate more research and development in the growing fields. Finally, the paper aims to act as a reference for academics performing research, industry professionals improving their processes, and students looking to understand the concept of machine learning in biosynthesis.

**Introduction**
In 1944, the field of microbial biosynthesis was first established industrially, with the antibiotic penicillin being mass produced by a fungi belonging to the Penicillium genus.[1] Even decades ago, the importance of microbial biosynthesis was apparent, with penicillin alone being responsible for boosting the life expectancy by around 20 years.[2] It also saved countless lives during World Wars, rendering infections curable.[3] Since the 1940s, biosynthesis has been completely modernized, thanks to the exponential improvement of technology. These days, its applications range from fertilizers to metal nanoparticles, making it applicable in virtually every field. Prominent applications include environmentally sustainable fertilizers, which can boost food production [4], and carbon-neutral synthesis of ethanol (an alternative energy for diesel) to reduce greenhouse gas emissions.[5]

In recent years machine learning (ML) has experienced an exponential increase in popularity and its use in chemistry and biology depicts a similar trend. Modern synthetic biology researchers, protein-synthesis specialists and metabolic engineering scientists have partially adopted machine learning to enhance the outputs of their experiments, using it in numerous different ways.[6] Due to the surge in processing power, available datasets and research into ML, its applications have spread into microbial biosynthesis. Its uses include regressive analysis of optimal conditions for best results, inputs and involved organisms to predict plausible outcomes [7], 3d-imaging of inputted protein codes or predicting synthesized molecules [8] and even predicting the effects of genetically engineering an organism. Despite the increase in research done involving ML in synthetic biology, we have barely scratched the surface, with new applications being discovered regularly, each promising improved results. Finally, different techniques in ML enable different applications of it, from identifying faults in a system to modeling the proteins produced by organisms, and how the organisms respond to external conditions. [9]

This review paper aims to provide a comprehensive overview of the intersection between machine learning and microbial biosynthesis, and how the application of different ML algorithms can benefit microbial biosynthesis. We compare ML-based methods to those that are traditional and provide possible recommendations for future research as well as current challenges.

**Basics of Microbial Biosynthesis**
Biosynthesis is defined as the process in which substrates are converted into more complex products by living organisms in a multi-step, enzyme-controlled pathway.[10] In an enzymatic pathway, a substrate undergoes successive catalytic transformations by enzymes, leading to the sequential release of intermediate products until the final product is synthesized. Microbial biosynthesis refers to the biologically mediated synthesis of molecules by microorganisms, such as bacteria and fungi. These unicellular organisms, capable of fermentation reactions, possess the remarkable capacity to synthesize and excrete complex molecules within their cellular structures. Furthermore, they generally have fast rates of reproduction, allowing a small sample to multiply into one with a significant number of the organisms. As such, instigating genetic changes within a smaller number of cells is feasible, making the process of fermentation cheaper and faster.[11] Finally, the utilization of these microorganisms does not entail ethical considerations, in contrast to the ethical considerations involved in experiments or research involving animals or humans.  These organisms can be used for biosynthesis due to the presence of plasmids, or small loops of DNA, which can be extracted and reinserted after introducing a strand of human DNA into it, in a process known as recombinant gene editing.[12]

For microbial biosynthesis, first a sample of the microorganisms is developed, usually using genetic alteration to include necessary genes to synthesize the accurate compounds. These genes are either taken from organisms that naturally produced said compounds or are artificially made in a laboratory. Genetic modification for biosynthesis is generally via combinatorial modification. During this procedure, the plasmids of the microorganisms are initially extracted, followed by the utilization of restriction enzymes to cleave them. Subsequently, the same enzymes are employed to excise the coding region from the human nucleus. The resulting fragments are then joined together using DNA ligase, which utilizes the complementary sticky ends of the strands, resulting in the formation of a recombinant plasmid harboring the human DNA. Finally, this recombinant plasmid is reintroduced into the microorganism.[13] The inserted genes code for a specific enzyme pathway that produces the desired product. These recombinant bacteria cells are duplicated in a nutrient medium to increase the sample size, after which industrial fermenters, which contain a nutrient medium optimal for survival, are inoculated with the organisms. Substrates are introduced from the top, and oxygen is supplied from below in the fermenters. Large stirrers are employed to ensure uniform substrate concentrations, keep the sample suspended, and regulate temperatures effectively. Additionally, these fermenters are equipped with sensors to monitor and maintain optimal parameters such as temperature, pH, and oxygen concentration.[14] The nutrient medium typically comprises sugars, ammonium ions, oxygen, and other essential minerals, including phosphates. Microorganisms utilize this medium, along with modified enzyme pathways, to synthesize the desired molecules. The synthesized molecules then diffuse into the nutrient medium until they are gradually removed.

Within the fields of medicine and biology, biosynthesis has numerous applications. For example, the synthesis of drugs is a major use, with over 300 biopharmaceutical products available. One such product utilizes E. coli cells to produce insulin on a large scale via genetic alteration. This is important for diabetics as a necessary supplement to offset their naturally low levels.[15] Another class of molecules that are extremely important is flavones. which are naturally produced within plants, and have uses like providing protection against UV radiation, pathogen resistance, and the development of the plants. In human diets, they play a massively different role, having health benefits including anti-inflammatory, anti-bacterial and cancer reducing properties.[16] As such, the increased synthesis of these compounds for human consumption would be greatly beneficial.

Another application is the synthesis of renewable, clean fuel sources. Bioethanol and biodiesel are two examples of this that have already made inroads into replacing petroleum-based fuels. These are environmentally friendly, carbon neutral, and made from plants and biowaste. Ethanol is made by fermenting glucose using yeast, and biodiesel from rapeseed. Biofuels can be made via direct photosynthesis using sunlight and $CO_2$ in photobioreactors and catalyzed conversion of biomass from farms using microorganisms.[17] This application could reduce carbon emissions, and the subsequent greenhouse effect greatly. Lastly, a final application would be in the synthesis of metal nanoparticles, notable for their use in material, electronics and energy industries, such as photonics [18]. Biosynthesis offers several advantages over chemical methods for nanoparticle synthesis. Firstly, it allows for the use of lower temperatures, which reduces energy consumption during the process. Secondly, biosynthesis is more environmentally friendly as it does not require the use of heavy metals, which are often utilized in chemical methods. Finally, this biological approach enables the synthesis of a broader spectrum of nanoparticles, expanding the possibilities for diverse applications. This synthesis is done via producing metal-binding proteins, within genetically modified E. coli cells that overexpress the metallothionein and phytochelatin peptides when exposed to heavy metal ions.[19]

The conventional application of combinatorial biosynthesis poses several challenges. One significant issue is the lack of specificity in pathway enzymes, resulting in the production of products that closely resemble the desired compound but differ in certain aspects. Consequently, the imperfect pathways lead to reduced yields compared to the wild variants.[20] Moreover, when attempting to incorporate additional products into the enzyme pathways, further complexities arise. Some desired products are not naturally produced by organisms, so altering enzyme pathways is necessary to produce this - a task that is extremely difficult due to the difficulty present in predicting the shape of the enzyme and resultant effect on the substrate. An example of this is altering the specificity of tailor enzymes for antibiotic maturation, to alter the chemical changes (methylation, glycosylation, oxidation etc).[21]

**Introduction to Machine Learning**

The field of machine learning is developing rapidly, with new techniques and programs being created daily. In the recent decade, ML has experienced massive improvements, for several key reasons. First, the availability of cheaper technology has enabled the development of devices with considerably faster processing speeds, allowing more data to be processed per second, and thus, more efficient models. Next, larger datasets being publicly available, due to more research done into fields and the expanding use of the internet, enables programs to be trained easily and accurately. Furthermore, as technology has developed, manufacturers of chips and CPUs are capable of clustering semiconductors/transistors at high densities, due to them being exceptionally small. As such, more processes (one program done [22]) can be done per second, and more data stored. Recently, deep learning has also been explored, using multiple layers of Artificial Neural Networks (programs designed after a human brain [23]), giving a powerful tool for regression, predictions, and recognition.[24].Thus, with recent advances in research, technology and datasets, Machine Learning is at its all-time high, and is expected to continue rising. As a result of this improvement in ML, there has been a corresponding rise in its applications in physics, engineering (for stable structures), autonomous cars and predictive analysis.

There are four major types of machine learning techniques (shown in Fig 1), supervised, unsupervised and semi-supervised learning, and reinforcement learning. In supervised learning, the model creates a function that derives desired outputs from the inputs by constructing a function to map inputs to desired outputs. The program is trained using labeled data, which provides definitive and accurate information, enabling the program to identify patterns effectively. Supervised learning has countless uses, one of which is diagnosing diseases. It is majorly done by using deep learning and convolutional neural networks (a subset of deep learning). Diagnosing diseases using ML has widespread applications, including dangerous ailments including cancers, or simple issues like skin diseases. Most of these programs use images and identify visible symptoms from these. Images of the skin diseases can be taken and uploaded to a ML program that can identify (within a certain probability) the disease, whether it's cancer and if its malignant, benign etc. A traditional ML model would separate the features from the image and then identify the skin disease. It is trained with labelled data (supervised learning) to learn how to separate the features, usually based on a color difference, size difference or texture difference with the background [25], and learns aspects of each type of skin disease. This sample data is labelled, to assign a specific disease to each pattern identified. Deep learning, however, would first clean the image, removing hair and other noise, then adjust the size/color and then diagnose it, getting a high-quality image due to preprocessing, allowing considerably higher success rates. Both use supervised learning, as it is better with multi-dimensions and continuous features, and work far better than unsupervised learning with a large dataset as is present for skin diseases.[26]

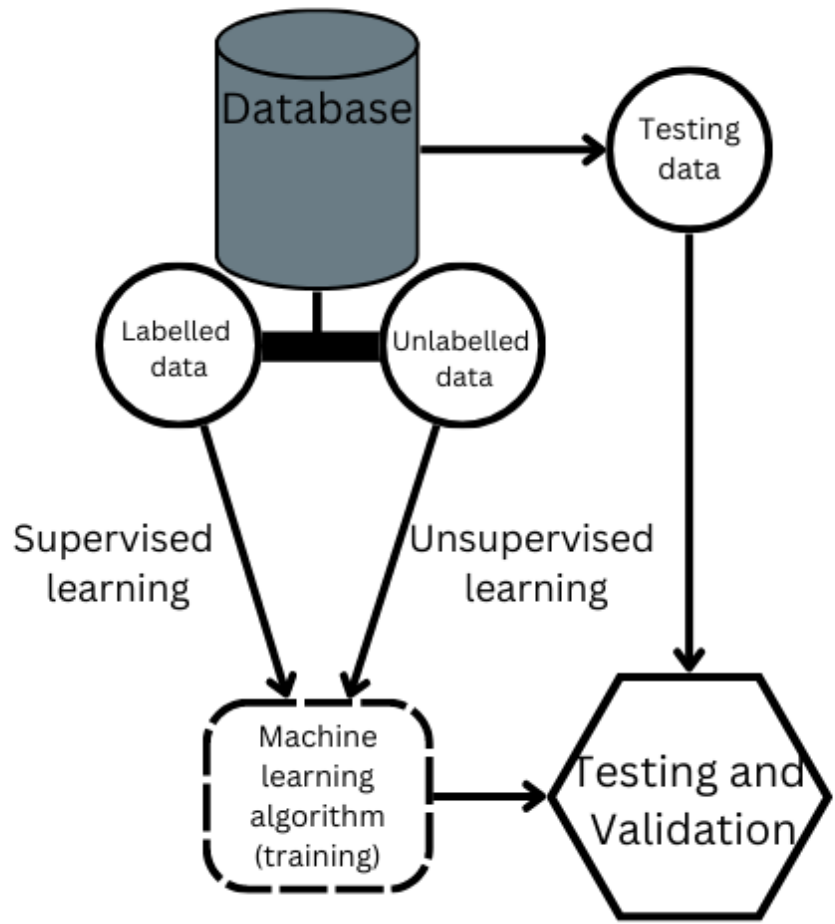

*Fig 1- description of Machine Learning program training types [44]*

On the other hand, unsupervised learning involves using unlabeled data, which is then clustered on a graph to identify similarities and discover trends. Here, the algorithm's ability to detect similarity becomes crucial, and it is extremely capable at detecting outliers in particular datasets, undefined characteristics of data and similarities within elements of data points. Alternatively, unsupervised learning is more capable at identifying trends or outliers in the data that is inputted, without a predefined output. As such, unsupervised learning is more applicable to remove or identify anomalies. One example of its use is to detect epileptic seizures by detecting epileptiform discharge. Epilepsy is a disorder which is characterized by repeated seizures. Generally, electric activity in the brain is detected and recorded by electroencephalography (EEG). Unsupervised learning trains algorithms that identify the instance of the seizure/s from the background EEG recording, enabling automated detection of the seizure.[27] Reinforcement learning programs have been used on CT (Computed Tomography), ultrasound and MRI scans, to identify individual bones/features of the image (for example, label the bones in the abdomen or the vocal structure).[28] This would make identifying fractures, and sometimes even cancer, considerably easier as the bulk of the features could be labelled automatically, leaving less for doctors to look at.

Semi-supervised learning combines aspects of both approaches, as it involves a mix of labeled and unlabeled datapoints and it's used for either task, especially when the dataset lacks uniformity.

Reinforcement learning involves the program performing a process/action, and receiving feedback based off it. The feedback is based on the rules the program follows [29], and reaffirms the pattern recognized/the rule.

Recently, deep learning has been given lots of attention due to its advanced capabilities in prediction and ability to compare different data types and datasets. Deep learning involves the construction of artificial neural networks with multiple layers, enabling the model to learn effectively from complex input data. Each layer, whether linear or non-linear, performs a specific function on the data and passes the processed information to the subsequent layer. The primary objective is to learn from the data and classify it based on weightage assigned to each layer. By leveraging this hierarchical architecture, deep learning models can effectively capture intricate patterns and relationships within the data, making them suitable for a wide range of tasks, including image and speech recognition, natural language processing, and many other complex problems. [25] Deep learning models are particularly good at computing massive datasets. They are applied in numerous fields, including medical diagnoses, autonomous cars, and image processing. In autonomous cars, it is used firstly for image recognition, in which it identifies objects recorded by cameras, as well as their distances, to create a map of the road ahead- allowing it to drive safely without crashing. Image processing is done by extracting individual features, and then classifying them based on features similar to stored ones.[30]

**Application of machine learning in microbial biosynthesis-**

Up to date, ML applications in biosynthesis have had a wide range, with numerous types of programs, and outcomes being recorded. There are three major pathways followed by biological researchers (shown in Fig 2): three-dimensional modelling of proteins and the products of the synthetic reactions, regressive modelling to determine the optimal conditions for the organism and designing optimal enzyme pathways to obtain the product from the substrate. Each of these pathways utilize different machine learning programs that are designed for these.

First, modelling of genes, their protein products/enzymes and the resultant products involves identifying the group of genes responsible for a specific product. This is done by identifying biosynthetic gene clusters- a group of genes close to each other that code the enzyme pathway for a product. Using ML genomic mining approaches helps understand natural product chemistry and identify genes involved in the synthesis of molecules. Furthermore, the structure of the natural product can be identified using the genomic sequence. Tools like antiSMASH and PRISM use different curated rules to determine structural scaffolds of the products from BGCs. Other methods include looking at a single BGC class and predicting the structure from precursors. Finally, the structure is elucidated using experimental methods, like purification to make up for the complexity of prediction.[8]

Optimizing external and internal conditions for the biosynthetic process is crucial to obtaining optimal product synthesis. An example is the optimization of enzyme concentrations for the rate of molecule turnover from a metabolic pathway, better known as flux. One study in which this is done is in ref. [7], where a new model, the GC-ANN or glass ceiling artificial neural network, is used. Data concerning enzyme concentrations and corresponding flux is input into both an Artificial Neural Network (ANN) model and a classification model. The experiment unfolds in three stages: preparation, execution, and validation. In the preparation stage, the classification model determines a high-flux rule (>12 μM/s) by utilizing Principal Component Analysis (PCA), reducing the extensive dataset into a more manageable size while preserving significant information, and constructing a neural network model that predicts flux based on enzyme concentration [31]. Flux values are categorized into five groups, with the high-flux rule founded on discriminant analysis and a decision tree. The execution stage involves generating new enzyme concentrations in compliance with the high-flux rule. These concentrations are inputted into the ANN, which then forecasts the flux. In the validation stage, the predicted flux undergoes verification via experimental or simulated means. Overall, the experiment had promising results, with flux improvements of 63%, and the assay's cost decrease of up to 25%.[7] Thus, modelling the optimal internal conditions directly causes improved efficiency of the system. Furthermore, modelling the external conditions is also important. As shown in ref. [32], manipulating the culture conditions can increase yield for large-scale synthesis of melanin. This can be optimized using regressive models or ANN models that can factor in features including the concentration of minerals, glucose, oxygen etc.

Designing the optimal enzyme pathway is a crucial element of bioengineering for biosynthesis, and currently uses two programs- ART (Automated Recommendation Tool) and PCAP (Principal Component Analysis of Proteomics). ART utilizes machine learning and probabilistic modelling techniques, to provide a set of recommended strains (of enzymes/proteins) to be built using sample-based optimization. ART does not employ deep learning due to the scarcity of available datasets.[33] ART streamlines the Design-Build-Test-Learn (DBTL) cycles integral to genetic engineering. Initially, a new pathway or protein is designed and built using DNA. Its efficiency is subsequently tested, and insights are garnered from the results. These learnings are then applied to the next DBTL cycle to develop an improved strain. ART specifically enhances the learning stage, as it trains with experimental data and can perform iterative DBTL cycles. Moreover, it can suggest an optimal strain or pathway based on minimal data, a critical feature given the dearth of extensive datasets in biosynthesis. PCAP results in similar outcomes, through different methods, with PCAP using principal component analysis to suggest new designs. This is shown in the experiment in ref. [33], where the production of limonene is analyzed. Limonene can be converted into pharmaceutical chemicals or hydrogenated into jet-fuel, making it an important resource. Its synthesis using E. coli, using a pathway consisting of 7 genes in the Mevalonate pathway, to which two genes are added, replacing the previous method of obtaining it from plant biomass. Within the experiment, 27 different versions of the one pathway were built, each varying in the promoters, induction time and induction strength. For each pathway, data about the limonene production and protein expression of the 9 genes involved was collected, and inputted into PCAP, which then recommended new designs, which were then built. The new strains yielded a 40% increase in limonene production and a 200% increase in bisabolene production (a product of the same pathway). Subsequently, the researchers replicated the experiment employing ART, which echoed the outcomes of the PCAP approach, generating the same three designs. However, ART conferred the added advantage of being an automated system, requiring no human intervention. [33]

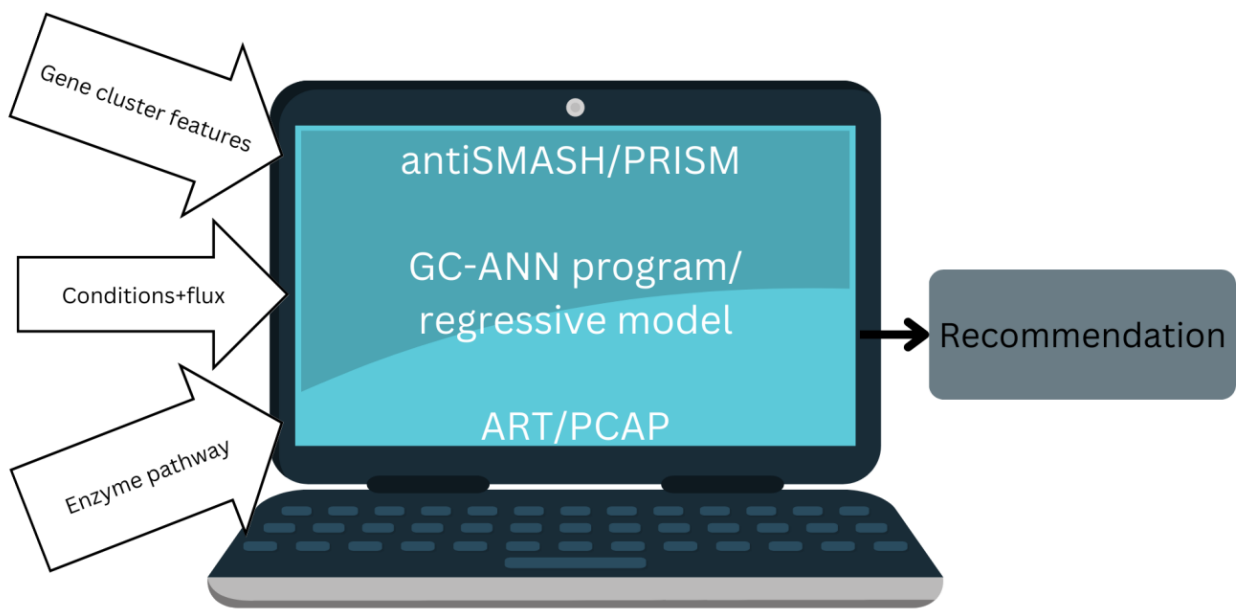

*Fig 2- Input data and output of ML programs in biosynthesis [44]*

Machine learning has numerous advantages and uses in biosynthesis, as summarized in table 1, but also several shortcomings. Firstly, ML can act as a substitute for expensive and time-consuming human labor, performing the same tasks faster and automatically, leaving the researchers time for practical experiments. The programs also perform these tasks more efficiently, e.g., the ART program. Second, ML can derive connections and identify subtle patterns and correlations within the data that humans might miss and can even omit irrelevant data that humans might mistake for a connection, as is done in PCA programs, giving smaller, more relevant datasets.[34] Furthermore, ML can process considerably larger datasets than humans can analyze, enabling more variables to be considered, especially when differing data types are used. Finally, certain programs can perform tasks like three-dimensional modelling, identifying structures of products based on genes, and identifying specific genes- all of which would be difficult for humans due to the innumerable possibilities [33]. Thus, automated programs are more suited to anything related to structural predictions or looking at datasets as big as the genetic code. Despite all these advantages, the application of ML is held back by certain shortcomings. For one, within microbial biosynthesis, datasets are generally extremely small, consisting of very few samples, leading to weakly trained models, and the inability to use ANN (not enough data to train the network), as shown in ref. [33], where a dataset of only 27 points was used. Furthermore, the lack of trained personnel who can develop/alter ML programs to be better suited for biosynthesis results in the use of technology that could be improved. Lastly, data can be overfit, meaning that it identifies a correlation/pattern that does not exist/is not relevant, resulting in inaccurate results [35]. In biosynthesis this could lead to designing enzymes that are inefficient, or produce incorrect products, leading to another issue. Some molecules are extremely similar chemically, simply having a small difference in the location of an arm of the molecule- causing drastically different potency. These might be produced because of an incorrect/overfit model, causing possibly dangerous results.[36]

| Characteristics | Traditional (recombinant plasmids) | BGC cluster detection (antiSMASH/PRISM) | GC-ANN | ART/PCAP |
|---|---|---|---|---|
| How it works | Introduces the relevant enzyme pathway, similar to as is found in a host. | Different rules are used as well as the process and product of each enzyme reaction to map the structure from previous data | Artificial neural networks use hidden layers to apply different weightages and create a model linking the output to input, and then identify the peak (layers and weightages depend on trends identified) | Use principal component analysis to make patterns clearer, and programs using inputted data to recommend changes to improve the functioning of the pathway by boosting the DBTL cycles |
| Method | The gene is inserted into a plasmid from another organism (no program used) | Genomic mining identifies groups of genes coding for a product and antiSMASH/PRISM generate a likely structure based on the BGCs. | Data (conditions + output) is inputted, which is then classified into sections to identify the general rule followed by the best results. The ideal conditions and their associated result are calculated and outputted by the ANN model | Input data of different enzymes and pathways for the product, and their effects, and build the recommendations |

| Use/efficacy | Used to synthesize molecules like insulin, that naturally occur in other organisms. Lower efficiency than if improved by ML | Allows products to be mapped to genes, enabling the genes to be transferred or identify areas where changes can be made for desired results. Low accuracy due to mapping not being fully developed | Used to identify the optimal internal and external conditions including enzyme conc. Very efficient, above 50% improvement. Better than the traditional method due to the conditions being specifically suited to improve the production. | Used to optimize pathways. Very effective, with 40-200% improvements. |
|---|---|---|---|---|

*Table 1- Machine Learning applications in biosynthesis*

**Recent Advances and Future Prospects**

Over recent years the intersection of the fields of microbial biosynthesis and machine learning has expanded, with numerous papers being published daily using it. In the late 20$^{th}$ century, the field was completely non-existent and has since risen to being a leading subpart of the field of microbial biosynthesis. This improvement is due to the improved processing power and capabilities of machine learning, larger datasets, and increased interest in the field of biosynthesis- leading to a deeper understanding of the field. Between 1990 and 2000, there were few papers published on the topic, approximately 350 publications selected by Google Scholar[37] from a search using the keywords ``Machine learning" and "biosynthesis", while the number between 2010 and 2020 is noticeably greater, at 14200.[38] Comparatively, the number between 2000 and 2010 is 3220[39]- indicating an increasing exponential curve for the number of publications of machine learning in biosynthesis. Since the difference between 2000-2010 and 1990-2000 is only 2870, and between 2000-2010 and 2010-2020 it is 10980, an upward trend of publications till the present is visible. This would increase the quality of datasets available for this field, as more researchers could provide data, explaining recent advances in it. Another indicator of the current boost in the application of this intersection, as well as the use's complexity, is the fact that the ref.[40] published in 2001 states that an approximate of 30 microbial genomes had been fully sequenced, while by June 2020 approximately 130 thousand microbial genomes had been sequenced.[41] This, once again, shows the exponential increase in availability of datasets, this time in the form of genomic data, essential to understand enzyme pathways and thus use programs like ART. Additionally, the 2001 paper also details the use of a machine learning algorithm, namely a decision tree. While the program (C4.5), like ART, identified the effect of changing a gene for phenotypic expression, it could not recommend suitable new genes or quantify the likely success of an enzyme pathway. C4.5 could only analyze the input data and develop rules to identify the likely class of a gene based on the difference in growth between wild-types and mutants when exposed to the same conditions. As such, ART, a newer technology, is far more advanced than the older program- displaying the obvious advances in the applications of machine learning for biosynthesis being aided by the development of newer technologies, with improved capabilities. These technologies also widen the range of applications of ML within biosynthesis. In the past these uses included regressive analysis of conditions and classifying genes/identifying optimal genes or conditions based on different conditions, genes and their related phenotypic expressions being inputted. Now, however, 3D imaging of proteins is possible, using programs like Alpha Fold (the foremost program in modelling, using template free modelling (modelling without an existing template), according to the Critical Assessment of Structure Prediction (CASP)) [42]. Additionally, the development of ART and PCAP enables designing custom intracellular enzyme pathways, far quicker and more effectively than previously possible. Lastly, genomic mining has also improved substantially, with ML currently being able to identify gene clusters, and thus help map the genome of species more efficiently than before when done manually/with basic programs- a fact made evident by the factor by which the number of sequenced genomes has increased.

The future of machine learning in microbial biosynthesis i bright, as there are several issues to be resolved and improvements to be made. For one, the development of larger datasets, possibly even a universal dataset comprising of subparts with like data. The universal dataset would be an accumulation of all the data used in research and acquired from experiments, to be used for analysis by any program. To continue, this dataset could be automatically updated with newly acquired information- as well as follow a uniform format, enabling more data to be eligible for use for any program. Finally, the presence of this extensive dataset would enable the widespread use of ANNs, which require more datapoints than traditional models to be accurate but can be functional in analyzing multiple datatypes and perform multiple tasks at once, allowing faster and more holistic results.[43] Another prospect would be to develop new programs and technologies capable of analyzing new variables, larger datasets and altogether reduce the wastage of valuable human time. For example, an extension to the ART program could be made to analyze the genetic code/phenotypic expression and natural products synthesized by microorganisms. It could then recommend a species that is most suitable (genetically, internal condition wise, adaptation to external conditions etc) to be genetically engineered with a recommended enzyme pathway. Another example of a new program would be to link medical and biosynthesis-related programs, interconnecting the diagnoses and treatment of diseases with production of medical compounds. Linking biosynthetic molecules directly with the treatment of diseases is difficult, as it involves large amounts of data that is needed to first identify the substance that treats an ailment, then 3d structure it and identify/design possible enzyme pathways, and finally construct and test said pathways. However, in the future, perhaps with the universal dataset, it could be plausible since advanced technology could do all these tasks with relatively few input datapoints, as well as with higher levels of accuracy. These would undoubtedly need to be incredibly powerful and trained with complex rules or even other smaller programs like a large-scale ANN made of computer running programs. Furthermore, a shorter-term plan would be to boost the functionality of existing ML programs, rather than developing new models, providing new possibilities soon in the future. This could be done by simply developing new versions of certain models, like ART that can also predict the bi-products or optimal conditions, or a new version of

Alpha fold able to predict qualities of the synthesized substances or the species of origination. Thus, researchers should focus on developing new models of ML for biosynthesis (primarily focused on neural networks), collecting large amounts of data, and improving existing technologies with new data and techniques.

Current problems with ML in biosynthesis occur from issues with both Machine Learning and with biosynthesis (shown in Fig 3). While these problems do hinder current progress, they also leave possible windows of development in the future, when they are resolved. For instance, the lack of data and sufficiently large datasets is a major issue. Due to the relative newness of ML in biosynthesis there is not a lot of data that is usable for studies. As such, some models are not usable or are extremely inaccurate due to insufficient datapoints and insufficient training data. Tha lack of data also prevents neural networks from being used which is detrimental to the functionality of programs like the ART program. One solution, as is given above, would be to develop larger datasets, having all published researchers upload their accumulated data in a standard format, and increase the number of experiments performed by making the importance of the field more obvious. Another issue with ML models is the generally low accuracy for more complex tasks (like classification, 3d imaging etc) caused by a lack of data, weaker models which are over or under sensitive to certain aspects of the inputs, and sometimes the lack of trained personnel to utilize it optimally. This would make these models inadequate for use due to them giving false data (enzyme pathways that would be faulty or produce dangerous chemically similar chemicals). To solve this problem, several methods are viable. First, a detailed testing stage for each model to ensure that it has an accuracy above a certain point and its output does not lead to something harmful. Then, ensuring the use of larger datasets by merging several or performing primary research would reduce the chances of negative consequences.

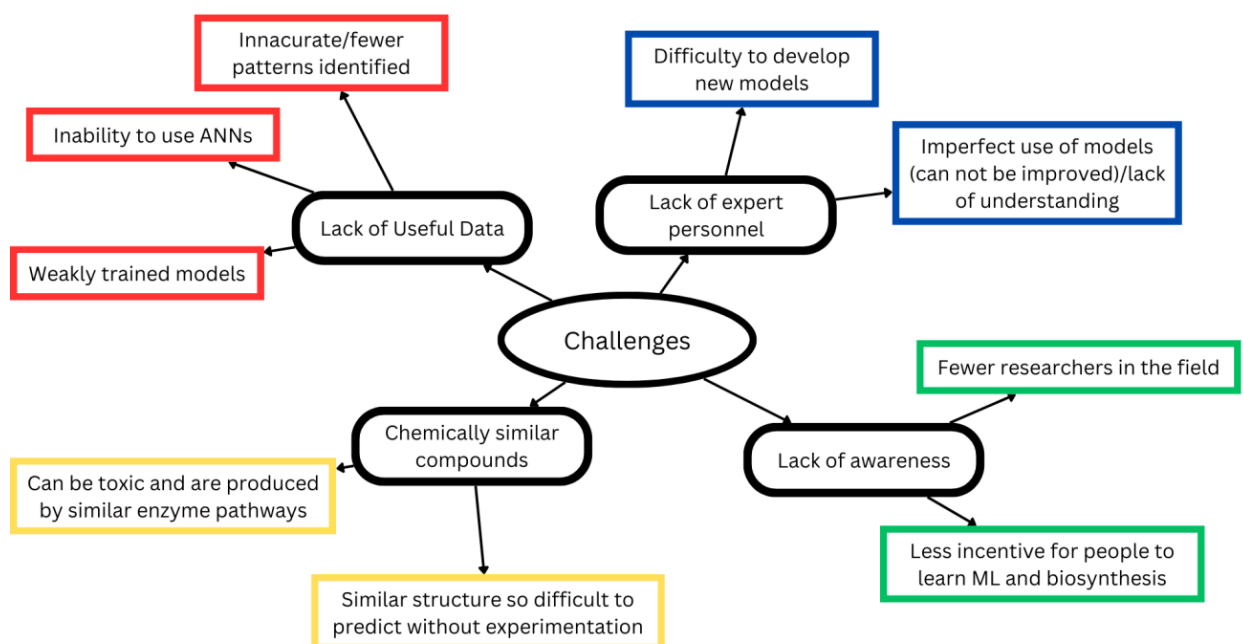

*Fig 3- Challenges of ML and biosynthesis [44]*

**Conclusion**

This paper has discussed the recent advances in ML and biosynthesis as well as potential future advances. The expanding application of machine learning in biosynthesis and biology is evident, following a historical trend of increasing adoption. This review highlights the primary advantages and methodologies associated with various machine learning programs for different applications. Machine learning offers distinct advantages over traditional methods, especially in handling extensive datasets and complex computational tasks. Its relevance in biosynthesis is significant, particularly in tasks where human expertise is limited. These advantages encompass faster data processing, increased accuracy, and the ability to identify patterns in vast datasets more effectively. The potential applications of machine learning are broad, ranging from improving animal health and production to facilitating more efficient industrial processes and aiding in genome mapping. Given the challenges and opportunities in the field, it's essential to prioritize data collection and standardize the recording of DBTL cycles. This data will be fundamental for future model development. The envisaged model, capable of identifying and characterizing synthetic molecules and their optimal conditions for biosynthesis, could have transformative impacts on industry, medical and food production, and scientific research. Therefore, professionals in microbial biosynthesis are encouraged to familiarize themselves with the capabilities of machine learning to further enhance the field's progress.

**List of Abbreviations**
ML - Machine Learning
DNA- Deoxyribonucleic acid
UV- Ultraviolet
$CO_2$- Carbon Dioxide
CPU- Central Processing Unit
EEG- Electroencephalography
CT- Computed Tomography
MRI- Magnetic Resonance Imaging
BGC- Biosynthetic Gene Clusters
ANN- Artificial Neural Network
GC-ANN- Glass ceiling- Artificial neural Network
PCA- Principal Component Analysis
ART- Automated Recommendation Technology
PCAP- Principal Component Analysis of Proteomics
DBTL- Design-Build-Test-Learn
CASP- Critical Assessment of Structure Prediction

**Declarations**

**Ethics approval and consent to participate.**
This study required no ethics approval and had no subjects who we needed to acquire consent from.

**Consent for publication**

Not applicable

**Availability of data and materials**

Data sharing is not applicable to this article as no datasets were generated or analyzed during the current study.

**Competing interests**

The authors declare that they have no competing interests.

**Funding**

The authors have no funding to declare.

**Authors' contributions**

AB and SR conceived the study. AB researched and wrote the manuscript. AB and SR edited the manuscript.

**Acknowledgements**

We would like to acknowledge Lumiere LLC for providing the platform to perform this research.